\documentclass[10pt]{article}
\usepackage{a4wide}
\usepackage{cite}

\newcommand{\dd}{{\rm{d}}} 
\newcommand{\rovno}{\!\!\!\!& = &\!\!\!\!}

\usepackage{color}

\begin{document}

\title{Gyratons in the Robinson--Trautman and Kundt classes}

\author{
Ji\v{r}\'{\i} Podolsk\'y
and
Robert \v{S}varc\thanks{
{\tt podolsky@mbox.troja.mff.cuni.cz}
and
{\tt robert.svarc@mff.cuni.cz}
}
\\ \ \\ \ \\
Institute of Theoretical Physics, Charles University, \\
V~Hole\v{s}ovi\v{c}k\'ach 2, 18000 Prague 8, Czech Republic.
}

\maketitle

\begin{abstract}
In our previous paper [Phys.~Rev.~D 89 (2014) 124029], cited as [1], we attempted to find Robinson--Trautman-type solutions of Einstein's equations representing gyratonic sources (matter field in the form of an aligned null fluid, or particles propagating with the speed of light, with an additional internal spin). Unfortunately, by making a mistake in our calculations, we came to the wrong conclusion that such solutions do not exist. We are now correcting this mistake. In fact, this allows us to explicitly find a new large family of gyratonic solutions in the Robinson--Trautman class of spacetimes in any dimension greater than (or equal to) three. Gyratons thus exist in all twist-free and shear-free geometries, that is both in the expanding Robinson--Trautman and in the non-expanding Kundt classes of spacetimes. We derive, summarize and compare explicit canonical metrics for all such spacetimes in arbitrary dimension.
\end{abstract}

\vfil\noindent
PACS class:  04.20.Jb, 04.30.--w, 04.50.--h, 04.40.Nr


\vfil
\eject

\section{Introduction}

Robinson--Trautman class of spacetimes \cite{RobTra60,RobTra62} together with the closely related Kundt class \cite{Kundt:1961} are important families of exact solutions to Einstein's field equations. They are geometrically defined by admitting a \emph{geodesic, shear-free and twist-free null congruence}. For the \emph{Robinson--Trautman} class such a congruence is \emph{expanding}, while for the \emph{Kundt} class it is \emph{non-expanding}.

In usual dimension ${D=4}$, these classes contain a great nuber of famous solutions, namely Schwarzschild-like static black holes, accelerating black holes ($C$-metric), Vaidya metric, Kinnersley photon rockets,  spacetimes with gravitational waves of various types (including well-known \emph{pp}-waves) propagating on various backgrounds (Minkowski, de~Sitter, anti-de~Sitter, direct-product universes etc.), and many other exact spacetimes. These are vacuum solutions with any value of the cosmological constant~$\Lambda$, they admit pure radiation, electromagnetic fields (both null and non-null), and other forms of matter. More details and specific references can be found, e.g., in chapters~28 and~31 of \cite{Stephani:2003} or chapters~18 and~19 of \cite{GriffithsPodolsky:2009}, respectively.

During the past decade, the large Robinson--Trautman class of solutions was extended to any higher dimension~${D>4}$ for the case of an empty space with any $\Lambda$ or aligned pure radiation~\cite{PodOrt06}, for aligned electromagnetic fields~\cite{OrtPodZof08}, and general p-form fields~\cite{OrtaggioPodolskyZofka:2015}. Similarly, extension of the Kundt class to higher dimensions was presented in~\cite{PodolskyZofka:2009}, see also \cite{ColeyFusterHervikPelavas:2006,ColeyHervikPapadopoulosPelavas:2009,OrtaggioPravdaPravdova:2013,SvarcPodolsky:2014b}. Complementarily, all Robinson--Trautman and Kundt solutions to Einstein's equations for $\Lambda$-vacuum, aligned pure radiation and gyratonic matter in lower dimension ${D=3}$ were recently found in \cite{PodolskySvarcMaeda:2018}.

\emph{Gyratonic matter} is a \emph{null field  with internal spin/helicity}. It was first considered already in 1970 by Bonnor \cite{Bonnor:1970b} who studied both the interior and the exterior solution of a ``spinning null fluid'' in the class of axially symmetric \emph{pp}-waves (see also Griffiths \cite{Griffiths:1972} who studied neutrino fields). Such matter is characterized not only by specific energy density profile, but also by non-zero angular momentum density profile. Spacetimes with localized spinning sources of this kind (spinning null particles accompanied by impulsive gravitational waves) moving at the speed of light were then independently rediscovered and investigated in 2005 by Frolov, Israel, Zelnikov and Fursaev \cite{FrolovFursaev:2005,FrolovIsraelZelnikov:2005}. These \emph{pp}-wave-type gyratons in ${D\ge4}$ were subsequently studied in greater detail, and also generalized to include ${\Lambda<0}$ \cite{FrolovZelnikov:2005}, electromagnetic field \cite{FrolovZelnikov:2006}, and various other settings including non-flat backgrounds or supergravity models. Summary of these gyratonic solutions can be found, e.g., in \cite{KrtousPodolskyZelnikovKadlecova:2012,PodolskySteinbauerSvarc:2012}.

All the so far known spacetimes with gyratonic matter sources belong to the Kundt class. Five years ago we asked ourselves a question: Are there gyratons in other geometries as well? The most natural candidate to investigate was the Robinson--Trautman class because it shares the twist-free and shear-free property. It differs only in having a non-vanishing expansion of the privileged null congruence. In our paper~\cite{SvarcPodolsky:2014} we attempted to systematically study the possible existence of Robinson--Trautman gyratonic solutions (in any dimension) which would be analogous to those known in the Kundt class. Unfortunately, by making a mistake in evaluating the gyratonic energy-momentum conservation equation, we came to the wrong conclusion that such solutions do not exist. Here we are correcting this specific mistake, and we explicitly derive a new large family of gyratonic solutions in the Robinson--Trautman class. \emph{Gyratons thus exist in all twist-free and shear-free ${D\ge3}$ geometries}.

In section~\ref{sec_geom} we summarize the general form of non-twisting shear-free geometries and Einstein's field equations, including the correct form of the gyratonic matter. Complete integration of  the field equations is presented in section~\ref{sec_inegration}. The obtained Robinson--Trautman spacetimes are summarized and discussed in concluding section~\ref{sec_discussion}. In particular, we compare the ${D>4}$, ${D=4}$, and ${D=3}$ cases. Moreover, in a compact and explicit form we present the entire class of Kundt solutions with aligned gyratonic matter in any dimension $D$, and we compare it with the newly obtained Robinson--Trautman class.


\section{General Robinson--Trautman and Kundt geometries and Einstein's equations for aligned gyratonic matter}
\label{sec_geom}

The \emph{metric} of the most general $D$-dimensional Robinson--Trautman or Kundt geometry can be written as
\begin{equation}
\dd s^2 = g_{pq}(r,u,x)\, \dd x^p\,\dd x^q+2\,g_{up}(r,u,x)\, \dd u\, \dd x^p -2\,\dd u\,\dd r+g_{uu}(r,u,x)\, \dd u^2 \,, \label{general nontwist}
\end{equation}
(see Eq.~(1) in~\cite{SvarcPodolsky:2014}) where $x$ is a shorthand for ${(D-2)}$ spatial coordinates ${x^p}$. Recall also that the nonvanishing contravariant metric components are $g^{pq}$ (an inverse matrix to $g_{pq}$), ${g^{ru}=-1}$, ${g^{rp}= g^{pq}g_{uq}}$ and ${g^{rr}= -g_{uu}+g^{pq}g_{up}g_{uq}}$ (so that ${g_{up}= g_{pq}g^{rq}}$ and ${g_{uu}= -g^{rr}+g_{pq}g^{rp}g^{rq}}$). The null vector field ${\mathbf{k}=\mathbf{\partial}_r}$ generates a geodesic and affinely parameterized null congruence which is twist-free and shear-free, provided ${g_{pq,r}=2\Theta\, g_{pq}}$. In the Robinson--Trautman class of geometries, this congruence has a nonvanishing expansion ${\Theta\ne0}$, while ${\Theta=0}$ defines the Kundt class.

Einstein's equations for the metric $g_{ab}$ read ${R_{ab}-\frac{1}{2}R\,g_{ab}+\Lambda\, g_{ab}=8\pi\, T_{ab}}$, where $\Lambda$ is any cosmological constant. We study spacetimes with a \emph{gyratonic matter aligned with} $\mathbf{k}$ \cite{Bonnor:1970b,FrolovFursaev:2005,KrtousPodolskyZelnikovKadlecova:2012}. In the coordinates of (\ref{general nontwist}), the nonvanishing components of the energy-momentum tensor $T_{ab}$ are
\begin{equation}
T_{uu}(r,u,x) \,, \qquad T_{up}(r,u,x) \,,
\end{equation}
where $T_{uu}$ corresponds to the classical pure radiation component while $T_{up}$ encode inner gyratonic angular momentum. Since its trace ${T\equiv g^{ab}\,T_{ab}}$ vanishes, Einstein's equations simplify to
\begin{equation}
R_{ab}= {\textstyle \frac{2}{D-2}\,\Lambda\,g_{ab}+8\pi\,T_{ab}} \,. \label{EinstinEq}
\end{equation}

In our previous paper \cite{SvarcPodolsky:2014}, we explicitly calculated all complicated components of the Ricci tensor $R_{ab}$, namely Eqs. (32)--(37). While these are correct, we made an \emph{unfortunate mistake in evaluating the conditions} ${T^{ab}_{\hspace{2.6mm};b}=0}$ following from the Bianchi identities. Indeed, Eqs. (54) and (55) in \cite{SvarcPodolsky:2014}  are wrong. Their correct form is
\begin{eqnarray}
&& T_{up,r}+(D-2)\,\Theta\, T_{up}=0 \,, \label{EqTup} \\
&& T_{uu,r}+(D-2)\,\Theta\, T_{uu}=g^{pq}\,T_{up||q}+g^{rp}_{\hspace{2.4mm},r}\,T_{up} \,, \label{EqTuu}
\end{eqnarray}
where the symbol ${\,_{||}}$ denotes the covariant derivative with respect to the spatial metric $g_{pq}$, that is
${T_{up||q} \equiv T_{up,q}-T_{um}\,^{S}\Gamma^{m}_{pq}}$ in which
${\,^{S}\Gamma^m_{pq}\equiv\frac{1}{2}g^{mn}(2g_{n(p,q)}-g_{pq,n})}$ are the Christoffel symbols with respect to the spatial coordinates only.

\section{Complete integration of the field equations}
\label{sec_inegration}

As in~\cite{SvarcPodolsky:2014}, we will now perform a step-by-step integration of the Einstein field equations (\ref{EinstinEq}) for ${\Theta\not=0}$. Some results will remain the same, but due to the corrected constrains (\ref{EqTup}), (\ref{EqTuu}), gyratonic solutions are actually found to exist.

\subsection{The equation ${R_{rr}= 0}$}
This field equation remains unchanged, providing us with the expansion scalar
\begin{equation}
\Theta=\frac{1}{r}\,, \label{ExplEx}
\end{equation}
and thus the ${(D-2)}$-dimensional spatial metric
\begin{equation}
g_{pq}=r^2\,h_{pq}(u,x) \,, \label{SpMetr}
\end{equation}
which are the same expressions as Eqs.~(57) and~(58) of~\cite{SvarcPodolsky:2014}.

\subsection{The equation ${R_{rp}= 0}$}
Also this equation has a correct solution given by Eqs.~(61) and~(62) of~\cite{SvarcPodolsky:2014}, that is
\begin{equation}
g^{rq}=e^{q}(u,x)+r^{1-D}f^{q}(u,x) \,, \label{NediagContra}
\end{equation}
and
\begin{equation}
g_{up}=r^2e_{p}(u,x)+r^{3-D}f_{p}(u,x) \,, \label{NediagCov}
\end{equation}
respectively. Here ${e_{p}\equiv h_{pq}e^q}$ and ${f_{p}\equiv h_{pq}f^q}$ are arbitrary functions of $u$ and $x$.

Using (\ref{ExplEx})--(\ref{NediagContra}), we can fully integrate the corrected energy-momentum conservation equations (\ref{EqTup}), (\ref{EqTuu}), yielding
\begin{eqnarray}
&& T_{up} = \mathcal{J}_p\,r^{2-D} \,, \label{ExplTup} \\
&& T_{uu} = \mathcal{N}\,r^{2-D}-{\mathcal{J}^p}_{||p}\,r^{1-D}+ f^p\mathcal{J}_p\,r^{3-2D} \,, \label{ExplTuu}
\end{eqnarray}
where ${\mathcal{J}_p(u,x)}$ and ${\mathcal{N}(u,x)}$ are arbitrary integration functions of $u$ and $x$, and ${{\mathcal{J}^p}_{||p}\equiv h^{pq}\mathcal{J}_{p||q}}$. These expressions rectify wrong Eqs.~(63) and (64) of~\cite{SvarcPodolsky:2014}.

\subsection{The equation ${R_{ru}= -\frac{2}{D-2}\,\Lambda}$}

Since this field equation is unaffected by the above-mentioned mistakes, Eq.~(67) of~\cite{SvarcPodolsky:2014} is correct, so that the corresponding metric function is
\begin{eqnarray}
g^{rr}\rovno a+b\,r^{3-D}+c\,r-\frac{2\Lambda}{(D-1)(D-2)}\,r^2  +\frac{D-3}{D-2}\,f^p\,\!_{||p}\,r^{2-D}+\frac{D-1}{2(D-2)}\,f^pf_p\,r^{2(2-D)} \,, \label{Explicit grr}
\end{eqnarray}
where
\begin{equation}
c\equiv -\frac{2}{D-2}\Big(e^{n}\,\!_{||n}-\frac{1}{2}h^{mn}h_{mn,u}\Big)\,, \label{c}
\end{equation}
which leads to
\begin{equation}
g_{uu}=-g^{rr}+r^2\,e^pe_p+2\,r^{3-D}\,e^pf_p+r^{2(2-D)}\,f^pf_p \,. \label{guuExpl}
\end{equation}

\subsection{The equation ${R_{pq}=\frac{2}{D-2}\,\Lambda\,g_{pq}}$}

This Einstein field equation was also correctly evaluated and integrated in~\cite{SvarcPodolsky:2014}. It turns out that in any dimension ${D \ge 4}$, necessarily
\begin{equation}
f_p=0  \label{fp0}
\end{equation}
for all ${(D-2)}$ spatial indices $p$ (interestingly, in \emph{lower} dimension ${D=3}$, the single function~$f$ remains arbitrary, see~\cite{PodolskySvarcMaeda:2018} and section~4.2 below). Consequently, the most general Robinson--Trautman line element takes the form
\begin{equation}
\dd s^2 = r^2\,h_{pq}\, \dd x^p\dd x^q+2\,r^2\,e_{p}\, \dd u \dd x^p -2\,\dd u\dd r+\big(r^2\,e^pe_p-g^{rr}\big)\, \dd u^2 \,,
\label{RTmetric}
\end{equation}
where
\begin{equation}
g^{rr}=a+b\,r^{3-D}+c\,r-\frac{2\Lambda}{(D-1)(D-2)}\,r^2 \,. \label{grr po Rpq}
\end{equation}
The functions ${h_{pq}}$ and $e_p$ are constrained by the equations
\begin{eqnarray}
\mathcal{R}_{pq}&\hspace{-2.0mm}=&\hspace{-2.0mm}\frac{\mathcal{R}}{D-2}\,h_{pq} \,, \label{EinSpace} \\
\frac{1}{2} h_{pq,u}&\hspace{-2.0mm}=&\hspace{-2.0mm}e_{(p||q)}+\frac{1}{2} \, c \, h_{pq}\,, \label{hpquDer}
\end{eqnarray}
that are also imposed by the field equation ${R_{pq}=\frac{2}{D-2}\,\Lambda\,g_{pq}}$, together with the relation
\begin{equation}
a = \frac{\mathcal{R}}{(D-2)(D-3)} \,. \label{aGrr}
\end{equation}
Here, ${\mathcal{R}\equiv h^{pq}\,\mathcal{R}_{pq}}$ is the Ricci scalar curvature of the spatial metric ${h_{pq}}$ which is the $r$-independent part of ${g_{pq}}$. Notice that due to (\ref{SpMetr}), the corresponding Ricci tensor is ${\mathcal{R}_{pq}\equiv \,^{S}R_{pq}}$, while ${\mathcal{R}\equiv \,^{S}R\,r^2}$.
Due to (\ref{EinSpace}), the transverse ${(D-2)}$-dimensional Riemannian space must be an Einstein space.

\subsection{The equation ${R_{up}=\frac{2}{D-2}\,\Lambda\,g_{up} +8\pi\,T_{up}}$}
This Einstein equation now takes the form
\begin{eqnarray}
&& {\textstyle -\frac{1}{D-2}\,\mathcal{R}\,e_p-\frac{D-3}{D-2}\big(e^{n}\,\!_{||n}-\frac{1}{2}h^{mn}h_{mn,u}\big)_{,p}+h^{mn}\big(h_{m[p,u||n]}+e_{[m,p]||n}\big)} \nonumber \\
&& {\textstyle +\frac{(D-4)}{2(D-2)(D-3)}\,\mathcal{R}_{,p}\,r^{-1}-\frac{1}{2}\,b_{,p}\,r^{2-D}} \nonumber \\
&& {\textstyle +\Big[(D-2)\big(e^ne_{[n,p]}-\frac{1}{2}(e^ne_n)_{,p}+\frac{1}{2}e^nh_{np,u}\big)+e_p\big(e^{n}\,\!_{||n}-\frac{1}{2}h^{mn}h_{mn,u}\big)\Big]\,r = 8\pi\,T_{up}} \,.
\end{eqnarray}
The gyratonic term $T_{up}$ on the right hand side is given by the corrected expression (\ref{ExplTup}), namely ${T_{up} = \mathcal{J}_p\,r^{2-D}}$. This gives us four conditions
\begin{eqnarray}
{\textstyle \mathcal{R}\,e_p+(D-3)\big(e^{n}\,\!_{||n}-\frac{1}{2}h^{mn}h_{mn,u}\big)_{,p}-(D-2)h^{mn}\big(h_{m[p,u||n]}+e_{[m,p]||n}\big)=0} \,, && \label{Rup r0} \\
(D-4)\,\mathcal{R}_{,p}=0 \,, && \label{Rup r-1} \\
b_{,p}=-16\pi\,\mathcal{J}_p  \,, && \label{Rup r2-D} \\
{\textstyle (D-2)\big(e^ne_{[n,p]}-\frac{1}{2}(e^ne_n)_{,p}+\frac{1}{2}e^nh_{np,u}\big)+e_p\big(e^{n}\,\!_{||n}-\frac{1}{2}h^{mn}h_{mn,u}\big) = 0 \label{Rup r} \,.} &&
\end{eqnarray}

In our previous paper we used \emph{wrong expression} ${T_{up}=\mathcal{J}_p\,r}$, which lead us to wrong relations ${b_{,p}=0}$ and subsequently ${\mathcal{J}_p=0}$, cf. Eqs.~(86) and (92) in \cite{SvarcPodolsky:2014}. Thus, we were mislead to the incorrect conclusion that there are no gyratonic solutions in the Robinson--Trautman class of geometries.
But such solutions do exist since nonzero $\mathcal{J}_p$ is obviously allowed by admitting a spatial dependence of the function $b(u,x)$ in (\ref{Rup r2-D}).

Moreover, as shown in our paper \cite{SvarcPodolsky:2014}, complicated equations (\ref{Rup r0}) and (\ref{Rup r}) are \emph{identically satisfied}. Equation (\ref{Rup r-1}) clearly restricts the dependence of the spatial Ricci scalar ${\mathcal{R}}$ on the spatial coordinates $x^p$, namely
\begin{eqnarray}
&& \mathcal{R}=\mathcal{R}(u)  \qquad \hbox{for} \quad D>4 \,, \label{R D>4}
\label{RDgr4}\\
&& \mathcal{R}=\mathcal{R}(u,x) \quad \hbox{for} \quad D=4 \,. \label{R D=4}
\end{eqnarray}
There is thus a \emph{significant difference} between the ${D=4}$ case of classical relativity and the extension of Robinson--Trautman spacetimes to higher dimensions. The remaining equation (\ref{Rup r2-D}) gives
\begin{equation}
\mathcal{J}_p=-{\textstyle \frac{1}{16\pi}}\,b_{,p} \,.
\label{Jpje0}
\end{equation}
Therefore, in \emph{any} dimension ${D \ge 4}$ we obtain the gyratonic matter component
\begin{equation}
T_{up} = -{\textstyle \frac{1}{16\pi}}\,b_{,p}\,r^{2-D} \,. \label{TuuExpl}
\end{equation}

\subsection{The equation ${R_{uu}=\frac{2}{D-2}\Lambda\,g_{uu} +8\pi\, T_{uu}}$}

This final equation determines the relation between the Robinson--Trautman geometry and the pure radiation matter field represented by the profile $\mathcal{N}(u,x)$ in (\ref{ExplTuu}).

For (\ref{ExplEx})--(\ref{NediagCov}) and (\ref{guuExpl}) with
(\ref{fp0}), the Ricci tensor component ${R_{uu}}$
becomes\footnote{Recall that ${e^{n}\,\!_{||n}\equiv
h^{nm}e_{m||n}}$, ${e_{p||q} \equiv
e_{p,q}-e_{m}\,^{S}\Gamma^{m}_{pq}}$, ${a_{||p||q} \equiv
a_{,pq}-a_{,n}\,^{S}\Gamma^{n}_{pq}}$ etc., see
\cite{SvarcPodolsky:2014} for more details.}
\begin{eqnarray}
&& R_{uu}= {\textstyle \frac{1}{2}g^{rr}g^{rr}_{\hspace{2.4mm},rr}+\frac{1}{2}\Big[e^{n}\,\!_{||n}-\frac{1}{2}h^{mn}h_{mn,u}+(D-2)\,g^{rr}\,r^{-1}-2\,e^ne_n\,r\Big]g^{rr}_{\hspace{2.4mm},r}} \nonumber\\
&& \hspace{10.0mm} {\textstyle +e^n\Big[g^{rr}_{\hspace{2.4mm},r}+\frac{1}{2}(D-6)\,g^{rr}r^{-1}\Big]_{,n}+\frac{1}{2}h^{mn}g^{rr}_{\hspace{2.4mm}||m||n}\,r^{-2}+\frac{1}{2}(D-2)\,g^{rr}_{\hspace{2.4mm},u}\,r^{-1}} \nonumber \\
&& \hspace{10.0mm} {\textstyle -(D-3)\,e^{n}e_n\,g^{rr}+h^{mn}\Big[e_{m,u||n}-\frac{1}{2}(e^pe_p)_{||m||n}-\frac{1}{2}h_{mn,uu}\Big]} \nonumber \\
&& \hspace{10.0mm} {\textstyle +h^{mn}h^{pq}\big(e_{[p,m]}+\frac{1}{2}h_{pm,u}\big)\big(e_{[q,n]}+\frac{1}{2}h_{qn,u}\big)} \nonumber \\
&& \hspace{10.0mm} {\textstyle +\Big[\frac{1}{2}(D-2)\big(e^me^nh_{mn,u}-e^n(e^pe_p)_{,n}\big)-e^pe_p\big(e^{n}\,\!_{||n}-\frac{1}{2}h^{mn}h_{mn,u}\big)\Big]\,r} \,.
\end{eqnarray}
Employing the explicit form (\ref{grr po Rpq}) of ${g^{rr}}$
with the help of (\ref{hpquDer}) we obtain
\begin{eqnarray}
&& R_{uu}= \textstyle{ \frac{2}{D-2}\,\Lambda\, g_{uu}}\nonumber\\
&& \hspace{11.0mm}\textstyle{
+\frac{1}{2}\Big[(D-2)b_{,u}+\frac{1}{2}(D-2)(D-1)\,b\,c-D\,e^{n}b_{,n}\Big]r^{2-D}}\nonumber\\
&& \hspace{11.0mm}\textstyle{
+\frac{1}{2}\triangle b\,r^{1-D}
+\frac{1}{2}\triangle a\,r^{-2}} \nonumber\\
&& \hspace{11.0mm} {\textstyle +\frac{1}{2}\Big[(D-2)\,(a_{,u}+a\,c)+(D-6)\,e^{n}a_{,n}
+\triangle c\Big]\,r^{-1}} \nonumber\\
&& \hspace{11.0mm} {\textstyle +\frac{1}{2}(D-2)\,(c_{,u}+c^2)+e^{n}\,\!_{||n}\,c
+\frac{1}{2}(D-4)\,e^n\,c_{,n}-(D-3)\,e^pe_p\,a} \nonumber\\
&& \hspace{18.0mm} {\textstyle +h^{mn}\Big[e_{m,u||n}-\frac{1}{2}h_{mn,uu}
-\frac{1}{2}(e^pe_p)_{||m||n}+h^{pq}e_{p||m}e_{q||n}\Big]} \nonumber\\
&& \hspace{11.0mm} {\textstyle +\frac{1}{2}(D-2)\,\Big[e^me^nh_{mn,u}-e^n(e^pe_p)_{,n}-e^ne_n\,c\Big]\,r} \,,
\label{Ruu1}
\end{eqnarray}
where $a$ is given by (\ref{aGrr}), $c$ is given by (\ref{c}),
and ${\triangle a \equiv h^{mn}a_{||m||n}}$ denotes the
covariant Laplace operator on the ${(D-2)}$-dimensional
transverse Riemannian space.

Now, in the Appendix of our previous work
\cite{SvarcPodolsky:2014} we proved the non-trivial identities
\begin{eqnarray}
e^me^nh_{mn,u}-e^n(e^pe_p)_{,n}-e^ne_n\,c=0 \,, \hspace{13.4mm}&&  \label{Identity1}\\
{\textstyle \frac{1}{2}(D-2)\,(c_{,u}+c^2)+e^{n}\,\!_{||n}\,c+\frac{1}{2}(D-4)\,e^n\,c_{,n}-(D-3)\,e^pe_p\,a}
\hspace{21.4mm} &&  \nonumber\\
{\textstyle +h^{mn}\big[e_{m,u||n}-\frac{1}{2}h_{mn,uu}-\frac{1}{2}(e^pe_p)_{||m||n}+h^{pq}e_{p||m}e_{q||n}\big]}=0
\,, \hspace{13.4mm}&&  \label{Identity2}\\
{\textstyle (D-2)\,(a_{,u}+a\,c)+(D-6)\,e^{n}a_{,n}+\triangle c}=(D-4)\,e^{n}a_{,n} \,, \hspace{-5.5mm}&& \label{Identity3}
\end{eqnarray}
which are valid in any dimension ${D \ge 4}$. These appear in
the terms in (\ref{Ruu1}) proportional to $r$, $r^0$, and
${r^{-1}}$, respectively. Einstein's equation
${R_{uu}=\frac{2}{D-2}\Lambda\,g_{uu} +8\pi\, T_{uu}}$ with
(\ref{ExplTuu}) thus simplifies to\footnote{Recall that
necessarily ${f^p=0}$, see (\ref{fp0}).}
\begin{eqnarray}
&&
\Big[(D-2)b_{,u}+\textstyle{\frac{1}{2}}(D-2)(D-1)\,b\,c-D\,e^{n}b_{,n}\Big]r^{2-D}
+\triangle b\,r^{1-D} \nonumber\\
&&\qquad
+\triangle a\,r^{-2}
+(D-4)\,e^{n}a_{,n}\,r^{-1}
= 16\pi\,\Big[\mathcal{N}\,r^{2-D}-{\mathcal{J}^p}_{||p}\,r^{1-D}\Big]\,.
\label{RTEq1}
\end{eqnarray}
Moreover, due to (\ref{Jpje0}) the gyratonic matter functions
${\mathcal{J}_p}$ always obey the ``divergence relation''
\begin{equation}
-16\pi\,{\mathcal{J}^p}_{||p}=\triangle b \,,
\label{Jp||q}
\end{equation}
so that the ${r^{1-D}}$ part of equation (\ref{RTEq1}) is
identically valid. Also, ${(D-4)\,a_{,n}=0}$ in any dimension
${D\ge4}$, see equations (\ref{Rup r-1}) and (\ref{aGrr}).
Consequently, the field equation (\ref{RTEq1}) reduces to
\begin{equation}
\Big[(D-2)b_{,u}+\textstyle{\frac{1}{2}}(D-2)(D-1)\,b\,c-D\,e^{n}b_{,n}\Big]r^{2-D}
+\triangle a\,r^{-2}=16\pi\,\mathcal{N}\,r^{2-D} \,. \label{RTEq}
\end{equation}
The factor ${\triangle a}$ proportional to ${r^{-2}}$ is always
zero in any ${D>4}$ due to (\ref{R D>4}), while in the ${D=4}$
case it is combined with the terms proportional to
${r^{2-D}=r^{-2}}$. The last Einstein's field equation thus reads
\begin{eqnarray}
{\textstyle
(D-2)b_{,u}+\frac{1}{2}}(D-2)(D-1)\,b\,c -D\,e^{n}b_{,n}=16\pi\,\mathcal{N}&&   \quad \hbox{for} \quad D>4 \,, \label{RTEq>4f}\\
\triangle ({\textstyle\frac{1}{2}}\mathcal{R})+2\,b_{,u}+3\,b\,c-4\,e^{n}b_{,n}=16\pi\,\mathcal{N}&&  \quad \hbox{for} \quad D=4 \,. \label{RTEq=4f}
\end{eqnarray}

This is a \emph{complete and explicit solution for gyratons
with aligned pure radiation in the Robinson--Trautman class of
geometries} (\ref{RTmetric}) in four and any higher dimension
$D$.

According to (\ref{Jpje0}), specific properties of the corresponding
gyraton are encoded in the metric function $b(u,x)$, and in the related
off-diagonal functions $e_p(u,x)$. The gyratonic matter is absent when
${\mathcal{J}_{p}=0}$, which is equivalent to ${b_{,p}=0}$.
In other words, there are no gyratons if (and
only if) the function $b(u)$ is independent of any
spatial coordinates.

\section{Summary and discussion}
\label{sec_discussion}

By fully integrating all Einstein's equations we explicitly proved that \emph{there are gyratons in the Robinson--Trautman class, as they are in the Kundt class}.
A null matter field in these geometries can thus have its ``internal spin''/angular momentum.

\subsection{Robinson--Trautman gyratons in ${D\ge4}$}

The most general $D$-dimensional (${D\ge4}$) Robinson--Trautman line element in vacuum, with a cosmological constant $\Lambda$, and possibly the pure radiation matter field with an additional gyratonic component, characterized by
\begin{eqnarray}
&& T_{up} = \mathcal{J}_p\,r^{2-D} \,, \label{sumExplTup} \\
&& T_{uu} = \mathcal{N}\,r^{2-D}-{\mathcal{J}^p}_{||p}\,r^{1-D} \,, \label{sumExplTuu}
\end{eqnarray}
can be written as
\begin{equation}
\dd s^2 = r^2h_{pq}\, \dd x^p\dd x^q
 +2r^2e_{p}\, \dd u\dd x^p
  -2\,\dd u\dd r+g_{uu}\, \dd u^2 \,, \label{RTmetricFin}
\end{equation}
where
\begin{equation}
g_{uu}=-\frac{\mathcal{R}}{(D-2)(D-3)}-\frac{b}{r^{D-3}}
+\frac{2}{D-2}\big(e^{n}\,\!_{||n}-{\textstyle \frac{1}{2}}h^{mn}h_{mn,u}\big)r
+\Big(\frac{2\Lambda}{(D-1)(D-2)}+e^n e_n\Big)r^2 \,,
\label{grrfinal}
\end{equation}
with the functions ${h_{pq}(u,x)}$, $e_p(u,x)$ and $b(u,x)$ constrained
by the field equations (\ref{EinSpace}), (\ref{hpquDer}) and (\ref{Rup r2-D}), (\ref{RTEq}), that is
\begin{eqnarray}
\mathcal{R}_{pq} \rovno \frac{h_{pq}}{D-2}\,\mathcal{R} \,, \label{EinSpacerep} \\
e_{(p||q)}-{\textstyle \frac{1}{2}}h_{pq,u} \rovno \frac{h_{pq}}{D-2}\,
\big(e^{n}\,\!_{||n}- {\textstyle \frac{1}{2}}h^{mn}h_{mn,u}\big)\,,  \label{hpquDerrep}
\end{eqnarray}
and
\begin{eqnarray}
-b_{,p} \rovno 16\pi\,\mathcal{J}_p   \,, \label{EinGyrat} \\
\frac{\triangle \mathcal{R}}{(D-2)(D-3)}-(D-1)
\big(e^{n}\,\!_{||n}-{\textstyle\frac{1}{2}}h^{mn}h_{mn,u}\big)\,b\,
+(D-2)\,b_{,u}  -D\,e^{n}b_{,n}\rovno 16\pi\,\mathcal{N} \,. \label{RTEqrep}
\end{eqnarray}

The first equation (\ref{EinSpacerep}) restricts the
Riemannian metric $h_{pq}$ of the transverse
${(D-2)}$-dimensional space covered by the coordinates~$x^p$
(with ${\mathcal{R}_{pq}}$ and ${\mathcal{R}}$ being its Ricci
tensor and Ricci scalar). \emph{Any Einstein space
metric} $h_{pq}$ \emph{is admitted}. The second constraint
(\ref{hpquDerrep}) imposes a \emph{specific coupling} between
this spatial metric $h_{pq}$ and the off-diagonal metric
components represented by ${(D-2)}$ functions ${e^p}$.

Equation (\ref{EinGyrat}) directly expresses the
gyratonic matter profile functions $\mathcal{J}_p(u,x)$ in (\ref{sumExplTup}) in terms of the
\emph{spatial derivatives of} $b(u,x)$ (recall also the relation (\ref{Jp||q}) which enables us to express the function ${\mathcal{J}^p}_{||p}$ in (\ref{sumExplTuu}) as ${ -\frac{1}{16\pi}\,\triangle b}$),
while equation (\ref{RTEqrep}) effectively relates these functions
to the pure radiation profile $\mathcal{N}(u,x)$.

In particular, \emph{in any higher dimension} ${D>4}$, the field equation (\ref{RTEqrep}) simplifies to
(\ref{RTEq>4f}), while \emph{in the usual} ${D=4}$ \emph{case} it takes the form (\ref{RTEq=4f}).
In the no-gyraton (${\mathcal{J}_p=0}$) case, that is for ${b_{,p}=0}$, equation (\ref{RTEq=4f})
reduces exactly to the classical Robinson--Trautman
equation (see \cite{Stephani:2003,GriffithsPodolsky:2009} with the
identification ${a={\textstyle\frac{1}{2}}\mathcal{R}=\triangle
(\log P)=K}$, ${b=-2m(u)}$,  ${c=-2(\log P)_{,u}}$, where $K$
is the Gaussian curvature of the spatial metric
${h_{pq}=P^{-2}\,\delta_{pq}}$). Equation (\ref{RTEq>4f})
generalizes the field equation previously derived in
\cite{PodOrt06} to admit the gyratonic matter in ${D>4}$.

\emph{Vacuum} spacetimes are obtained
when ${\mathcal{J}_p=0=\mathcal{N}}$. First of all, this  arises when ${b=0}$ (and  $\mathcal{R}$ is constant, which is true in any ${D>4}$ due to (\ref{Rup r-1})).

\subsection{Comparison to Robinson--Trautman gyratons in ${D=3}$}

In our recent work \cite{PodolskySvarcMaeda:2018}, we integrated Einstein's field equations for a general 3-dimensional Robinson--Trautman metric in vacuum, with a cosmological constant $\Lambda$, and possibly a pure radiation field and gyratons. The matter field takes the form
\begin{eqnarray}
&& T_{ux} = \frac{\mathcal{J}}{r} \,, \label{RTTup}\\
&& T_{uu} = \frac{\mathcal{N}}{r}-\frac{P(P\mathcal{J})_{,x}}{r^2}+\frac{fP^2\mathcal{J}}{r^3} \,, \label{RTTuu}
\end{eqnarray}
where ${\mathcal{N}(u,x)}$ and ${\mathcal{J}(u,x)}$ are functions determining the (density of) energy and angular momentum. The corresponding generic metric can be written in the form
\begin{eqnarray}
\dd s^2 \rovno \frac{r^2}{P^2}\, \dd x^2+2\,(e\,r^2+f\,)\,\dd u \dd x -2\,\dd u\dd r \nonumber \\
&& +\Big(-a  +2\big[ P(Pe)_{,x}+(\ln P)_{,u} \big]\,r +(\Lambda+P^2e^2)\,r^2\Big)\, \dd u^2 \,. \label{RTmetricsummary}
\end{eqnarray}
The  functions $P(u,x)$, $e(u,x)$, $f(u,x)$ and $a(u,x)$ are constrained just by two equations, namely
\begin{eqnarray}
 a_{,x} \rovno  c f - 2 f_{,u}-16\pi\, \mathcal{J} \,, \label{RTE1}\\
 a_{,u}\rovno ac+\triangle c+2(\Lambda+P^2e^2)P(Pf)_{,x} +3P^2f(P^2e^2)_{,x}  \nonumber\\
&& -2P^2f\,e_{,u} -P^2e(4 f_{,u}-cf+48\pi\, \mathcal{J})
+ 16\pi\,\mathcal{N} \,, \label{RTE2}
\end{eqnarray}
where ${\triangle c \equiv P(Pc_{,x})_{,x}}$
is the transverse-space Laplace operator applied on the function $c$, defined by $c \equiv 2\big[P(Pe)_{,x}+(\ln P)_{,u}\big]$.

Generically, by prescribing an \emph{arbitrary gyratonic function} $\mathcal{J}$ (as well as \emph{any} metric functions ${P, e, f}$) we can always integrate (\ref{RTE1}) to obtain ${a(u,x)}$. Subsequently, its partial derivative $a_{,u}$ (and other given functions) uniquely determines the pure radiation energy profile $\mathcal{N}$ via the field equation (\ref{RTE2}).

It is remarkable that in ${D=3}$ the function $f(u,x)$ in the metric (\ref{RTmetricsummary}) \emph{remains arbitrary} and, in general, \emph{non-vanishing}. This is an \emph{entirely new feature which does not occur in dimensions} ${D\ge4}$. Indeed, it was demonstrated in \cite{PodOrt06,OrtPodZof08,OrtaggioPodolskyZofka:2015} that for the Robinson--Trautman class of spacetimes in four and any higher dimensions necessarily ${f_p=0}$ for all ${(D-2)}$ spatial components. In this sense, the ${D=3}$ case is \emph{surprisingly richer} than the ${D\ge4}$ cases.

In the \emph{specific subcase} ${f=0}$, the metric (\ref{RTmetricsummary}) basically reduces to the form (\ref{RTmetricFin}), (\ref{grrfinal}) (where, of course, ${\mathcal{R}=0}$) with the two remaining field equations (\ref{RTE1}), (\ref{RTE2}) simplifying considerably to
\begin{eqnarray}
 a_{,x} \rovno  -16\pi\, \mathcal{J} \,, \label{RTE1f=01}\\
 a_{,u}\rovno ac+\triangle c-48\pi\,P^2e\, \mathcal{J}+ 16\pi\,\mathcal{N} \,. \label{RTE2f=0}
\end{eqnarray}
Since  $a$ here corresponds to $b$ in (\ref{grrfinal}), these two equations are  very similar to equations
(\ref{EinGyrat}), (\ref{RTEqrep}). The only difference is the additional term $\triangle c$ in (\ref{RTE2f=0}). In fact, it is not possible to  set ${D=3}$ in (\ref{RTEqrep}) because in this number of dimensions the terms in (\ref{Ruu1}) proportional to $r^{2-D}$ and $r^{-1}$ combine together, introducing thus the term $\triangle c$ into the correct field equation (\ref{RTE2f=0}).

\newpage

\subsection{Comparison to Kundt gyratons in ${D\ge3}$}

Finally, it is useful to compare the newly found complete class of Robinson--Trautman-type (${\Theta\not=0}$) gyratons in any dimension ${D}$ with the most general gyratonic solutions in the closely related Kundt family (${\Theta=0}$) of spacetimes, completing thus the derivation of \emph{all solutions with aligned gyratonic matter in any non-twisting and shear-free geometry}.

We obtain the most general Kundt gyratons by a direct integration of the field equations, using the explicit form of the Ricci tensor components which we presented in \cite{SvarcPodolsky:2014}. By setting ${\Theta=0}$, they simplify considerably. First, from the geometric relation ${g_{pq,r}=2\Theta\, g_{pq}}$ we immediately obtain ${g_{pq}=h_{pq}(u,x)}$ independent of $r$, instead of (\ref{SpMetr}) in the Robinson--Trautman case. The second field equation ${R_{rp}= 0}$ for ${\Theta=0}$ yields ${g_{up}=e_{p}+f_{p}\,r}$, so that ${g^{rp}=e^{p}+f^{p}\,r}$ (recall that ${e^p\equiv h^{pq}e_q}$, ${f^p\equiv h^{pq}f_q}$). The gyratonic/pure radiation matter field is then obtained by integrating (\ref{EqTup}), (\ref{EqTuu}) as
\begin{eqnarray}
&& T_{up} = \mathcal{J}_p \,, \label{KundtsumExplTup} \\
&& T_{uu} = \mathcal{N} +
({\mathcal{J}^p}_{||p}+ f^p\mathcal{J}_p)\,r \,, \label{KundtsumExplTuu}
\end{eqnarray}
where ${\mathcal{J}_p}$ and ${\mathcal{N}}$ are arbitrary functions of $u$ and $x$.
Einstein's equation ${R_{ru}= -\frac{2}{D-2}\,\Lambda}$ gives ${g_{uu}=a\,r^2+b\,r+c}$ with\footnote{The meanings of $a,b,c,e_p,f_p$ is here, of course, different from those in the Robinson--Trautman case.}
\begin{equation}
a=\frac{2\Lambda}{D-2}+{\textstyle \frac{1}{2}}(f^{p}\,\!_{||p}+f^p f_p)\,, \label{def_a}
\end{equation}
so that the Kundt metric takes the form
\begin{equation}
\dd s^2 = h_{pq}\, \dd x^p\dd x^q+2\,(e_{p}+f_{p}\,r)\,\dd u\dd x^p -2\,\dd u\dd r+(a\,r^2+b\,r+c)\, \dd u^2 \,. \label{KundtmetricFin}
\end{equation}
The next field equation ${R_{pq}=\frac{2}{D-2}\,\Lambda\,g_{pq}}$ yields just one constraint, namely
\begin{equation}
\mathcal{R}_{pq}= \frac{2\Lambda}{D-2}\,h_{pq}+f_{pq}\,,
\qquad\hbox{where}\qquad f_{pq}\equiv f_{(p||q)}+{\textstyle \frac{1}{2}} f_pf_q \,.\label{Rpq_constraint}
\end{equation}
It couples the Ricci curvature $\mathcal{R}_{pq}$ of the $(D-2)$-dimensional spatial metric $h_{pq}$ to the tensor $f_{pq}$ constructed from the functions $f_p$ determining the metric components $g_{up}$\,. The trace of (\ref{Rpq_constraint}) is ${\mathcal{R}= 2\,\Lambda+f^{p}\,\!_{||p}+{\textstyle \frac{1}{2}}f^p f_p}$, which enables us to rewrite $a$ as
\begin{equation}
a={\textstyle \frac{1}{2}} \mathcal{R}-\frac{D-4}{D-2}\,\Lambda+{\textstyle \frac{1}{4}}f^p f_p\,. \label{def_aALTER}
\end{equation}
Evaluating the field equation ${R_{up}=\frac{2}{D-2}\,\Lambda\,g_{up} +8\pi\,T_{up}}$, we obtain the following two conditions
\begin{eqnarray}
a_{,p}+{\textstyle\frac{1}{2}}f_p({f^n}_{||n}+f^nf_n)-2f^nf_{[n,p]}-h^{mn}f_{[m,p]||n}+\frac{2\Lambda}{D-2}\,f_p\rovno 0\,,\label{KundtEqrep}\\
b_{,p}-f_{p,u}-e^n(f_{n||p}-2f_{p||n}-f_pf_n)+f_p({e^n}_{||n}-{\textstyle\frac{1}{2}}h^{mn}h_{mn,u})&& \nonumber\\
-f^ne_{n||p} - 2h^{mn}(h_{m[p,u||n]}+e_{[m,p]||n}) +\frac{4\Lambda}{D-2}\,e_p\rovno -16\pi\,\mathcal{J}_p   \,. \label{EinGyratKUndt}
\end{eqnarray}
Effectively, they determine the spatial derivatives of the metric functions $a$ and $b$, respectively.
The last Einstein equation ${R_{uu}=\frac{2}{D-2}\Lambda\,g_{uu} +8\pi\, T_{uu}}$ contains terms proportional  $r^2$, $r^1$, and ${r^0}$. Separately, they form three constraints, namely
\begin{eqnarray}
&&\triangle a+{f^n}_{||n}\,a +3 f^n a_{,n} +2 f^n f_n\,a
-2h^{mn}h^{pq}f_{[p,m]}f_{[q,n]} = 0\,,\label{KundtRuur2}\\
&& \triangle b+f^n b_{,n}+4e^n a_{,n}
+2a ({e^n}_{||n}-{\textstyle\frac{1}{2}}h^{mn}h_{mn,u})+4f^ne_n\,a-2f^nf_{n,u}-4f^ne^m f_{[n,m]} \nonumber\\
&&\qquad - 2h^{mn}f_{m,u||n} - 2h^{mn}h^{pq}f_{[p,m]}(2e_{[q,n]}+h_{qn,u}) = -16\pi\,({\mathcal{J}^p}_{||p}+ f^p\mathcal{J}_p)  \,, \label{KundtRuur1}\\
&&\triangle c-{f^n}_{||n}\,c-f^n c_{,n}+2e^n b_{,n}
+b ({e^n}_{||n}-{\textstyle\frac{1}{2}}h^{mn}h_{mn,u})+h^{mn}h_{mn,uu} \nonumber\\
&&\qquad+2e^ne_n\,a-e^ne_nf^m f_m+e^nf_ne^m f_m-2e^nf_{n,u}-4f^ne^m e_{[n,m]}- 2h^{mn}e_{m,u||n} \nonumber\\ &&\qquad-2h^{mn}h^{pq}(e_{[p,m]}+{\textstyle\frac{1}{2}}h_{pm,u})(e_{[q,n]}+{\textstyle\frac{1}{2}}h_{qn,u}) = -16\pi\,\mathcal{N}  \,. \label{KundtRuur0}
\end{eqnarray}
Surprisingly, a lengthy calculation (using (\ref{def_a}), (\ref{Rpq_constraint}), standard properties of covariant derivatives, the identity (A.15) from \cite{SvarcPodolsky:2014}, and also the Bianchi identities) reveals that equations (\ref{KundtRuur2}) and (\ref{KundtRuur1}) are, in fact, \emph{identically satisfied} as a consequence of previous equations (\ref{KundtEqrep}) and (\ref{EinGyratKUndt}).\footnote{As shown previously in \cite{OrtaggioPravda:2016}, see footnote~8, the same is true for the Kundt spacetimes with aligned electromagnetic field.} We thus conclude that
the most general Kundt metric with aligned gyratonic matter can be written in the form (\ref{KundtmetricFin}) with (\ref{Rpq_constraint}), in which the metric function $a$ given by (\ref{def_aALTER}) is constrained by (\ref{KundtEqrep}), the function $b$ is determined by (\ref{EinGyratKUndt}), and the function $c$ satisfies equation (\ref{KundtRuur0}). The particular subcase ${D=3}$ is presented and discussed in more detail in \cite{PodolskySvarcMaeda:2018}.

There is a \emph{great simplification in the case when} ${f_p=0}$ for all $p$. In fact, it was shown in our previous work \cite{PodolskyZofka:2009} that this is a geometrically distinct subclass of the Kundt class. The complete family of such gyratonic solutions reads
\begin{equation}
\dd s^2 = h_{pq}\, \dd x^p\dd x^q+2\,e_{p}\,\dd u\dd x^p -2\,\dd u\dd r+\Big(\frac{2\Lambda}{D-2}\,r^2+b\,r+c\Big)\, \dd u^2 \,, \label{KundtmetricFinf=0}
\end{equation}
where, as in the Robinson--Trautman case, cf. (\ref{EinSpace}), $h_{pq}$ is the spatial metric of any Einstein space,
\begin{equation}
\mathcal{R}_{pq}= \frac{2\Lambda}{D-2}\,h_{pq}\,, \qquad \mathcal{R}= 2\Lambda \,,\label{Rpq_constraintf=0}
\end{equation}
equation (\ref{KundtEqrep}) is satisfied identically, and equations (\ref{EinGyratKUndt}), (\ref{KundtRuur0}) for the functions $b,c$ reduce to
\begin{eqnarray}
b_{,p}- 2h^{mn}(h_{m[p,u||n]}+e_{[m,p]||n}) +\frac{4\Lambda}{D-2}\,e_p\rovno -16\pi\,\mathcal{J}_p   \,, \label{Kundtf=0}\\
\triangle c+2e^n b_{,n}
+b ({e^n}_{||n}-{\textstyle\frac{1}{2}}h^{mn}h_{mn,u})+h^{mn}h_{mn,uu}+\frac{4\Lambda}{D-2}\,e^ne_n && \nonumber\\
- 2h^{mn}e_{m,u||n}- 2h^{mn}h^{pq}(e_{[p,m]}+{\textstyle\frac{1}{2}}h_{pm,u})(e_{[q,n]}+{\textstyle\frac{1}{2}}h_{qn,u})\rovno -16\pi\,\mathcal{N}  \,, \label{KundtRuur0f=0}
\end{eqnarray}
respectively. Equation (\ref{Kundtf=0}) relating $b_{,p}$ to $\mathcal{J}_p $ is similar to equation (\ref{Rup r2-D}) in the Robinson--Trautman case, while equation (\ref{KundtRuur0f=0}) relates the metric function $c$ to $\mathcal{N}$. The corresponding gyratonic matter takes the form
\begin{eqnarray}
&& T_{up} = \mathcal{J}_p \,, \label{KundtsumExplTupf=0} \\
&& T_{uu} = \mathcal{N} + {\mathcal{J}^p}_{||p}\,r \,, \label{KundtsumExplTuuf=0}
\end{eqnarray}
In fact, this ${f_p=0}$ subclass of Kundt spacetimes (\ref{KundtmetricFinf=0})--(\ref{KundtsumExplTuuf=0}) contains \emph{all} particular gyratonic solutions discussed in the literature so far, see~\cite{KrtousPodolskyZelnikovKadlecova:2012,PodolskySteinbauerSvarc:2012}  for a review and a list of references.

\section*{Acknowledgments}

We are very grateful to Hideki Maeda. During the joint work on paper \cite{PodolskySvarcMaeda:2018} concentrating on ${D=3}$, he pointed out our mistake in evaluating ${T^{ab}_{\hspace{2.6mm};b}=0}$ for the gyratonic matter. We immediately realized that we made the same mistake also in the ${D\ge4}$ cases, that is in our previous paper \cite{SvarcPodolsky:2014} which we are correcting here.
We acknowledge the support from the Albert Einstein Center for Gravitation and Astrophysics, Czech Science Foundation GACR~14-37086G.

\end{document}